
\headline={\ifnum\pageno=1\firstheadline\else
\ifodd\pageno\rightheadline \else\leftheadline\fi\fi}
\def\firstheadline{\hfil}
\def\rightheadline{\hfil}
\def\leftheadline{\hfil}
        \footline={\ifnum\pageno=1\firstfootline\else\otherfootline\fi}
\def\firstfootline{\rm\hss\folio\hss}
\def\otherfootline{\hfil}
\def\sqr#1#2{{\vbox{\hrule height.#2pt\hbox{\vrule width
.#2pt height#1pt \kern#1pt\vrule width.#2pt}\hrule height.#2pt}}}
\def\Box{\mathchoice\sqr64\sqr64\sqr{4.2}3\sqr33}
\def\rijkl{R^i{}_{jkl}}
\def\grijkl{\hat R^i{}_{jkl}}

\def\met {g_{\mu\nu}}

\font\tenbf=cmbx10
\font\tenrm=cmr10
\font\tenit=cmti10
\font\elevenbf=cmbx10 scaled\magstep 1
\font\elevenrm=cmr10 scaled\magstep 1
\font\elevenit=cmti10 scaled\magstep 1

\nopagenumbers
\line{\hfil }
\vglue 1cm
\hsize=6.0truein
\vsize=8.5truein
\parindent=3pc
\baselineskip=10pt
\centerline{\tenbf INSTANTONS, MONOPOLES, STRINGS AND FIVEBRANES}
\vglue 1.0cm
\centerline{\tenrm RAMZI R. KHURI}
\baselineskip=13pt
\centerline{\tenit Physics Department, Texas A\&M University}
\baselineskip=12pt
\centerline{\tenit College Station, TX 77843, USA}
\vglue 0.3cm
\centerline{\tenrm ABSTRACT}
\vglue 0.3cm
{\rightskip=3pc
 \leftskip=3pc
 \tenrm\baselineskip=12pt
 \noindent
We discuss three classes of solitonic solutions in string theory: instantons,
monopoles and string-like solitons. Instantons may provide a nonperturbative
understanding of the vacuum structure of string theory, while monopoles may
appear in string predictions for grand unification. The particular monopole
solution shown has finite action as the result of the cancellation of gauge and
gravitational singularities, a feature, which if it survives quantization,
may yield insight into the structure of string theory as a finite theory of
quantum gravity. In $D=10$, both monopoles and instantons possess fivebrane
structure. The string-like solitons represent extended states of fundamental
strings.{\it Talk given at INFN Eloisatron Project: 26th Workshop:
``From Superstrings to Supergravity'', Erice, Italy, Dec. 5-12, 1992.}
CTP/TAMU-82/92, December 1992.
\vglue 0.8cm }
\line{\elevenbf 1. Introduction \hfil}
\bigskip
\baselineskip=14pt
\elevenrm
In the past few years, classical solitonic solutions in string theory with
higher-membrane structure have been actively investigated. These solutions are
static multi-soliton solutions obeying a zero-force condition and saturating a
Bogomol'nyi bound between mass and charge. In certain cases, exact solutions of
bosonic and heterotic string theory may be constructed, each solution in
principle corresponding to an exact conformal field theory (CFT) of the
sigma-model. Although the solutions are initially conceived as perturbative
expansions in the classical string parameter $\alpha'$ (the inverse string
tension), the exact solutions acquire nonperturbative status. Being classical,
the solitons are tree-level solutions in the (quantum) topological expansion
of the string worldsheet, but are also nonperturbative in the loop parameter
$e^{\phi_0}$. Therefore, full quantum string-loop extensions of these
solutions await an understanding of nonperturbative string theory.
Nevertheless, it is possible to use these solitons in nonperturbative
calculations (such as vacuum-tunneling) since it is often the case that
higher order corrections do not contribute to these effects.

In this work we confine ourselves to three classes of solitons,
concentrating mainly on the construction of tree-level or exact static,
multi-solitonic solutions in bosonic or heterotic string theory$^1$. We
also briefly discuss the dynamics of these solitons$^2$. The three classes of
solitons we consider are: instanton solutions,
possessing instanton structure and local four-dimensional spherical symmetry,
monopole solutions, possessing magnetic monopole features and local
three-dimensional spherical symmetry, and ``cosmic string'' solutions, which
represent extended states of fundamental strings and correspond to
eight-dimensional instantons in $D=10$. Both the monopole and instanton
solutions have ``fivebrane''structure in $D=10$ (i.e. they are
$5+1$-dimensional objects in $9+1$-dimensional spacetime).

The motivation for the study of these solutions includes the following:
to begin with, these solitons represent solutions of string theory in
curved backgrounds, in contrast to the usual flat Minkowski background
solutions. Another interest in these
solutions stems from the application of known field-theoretic techniques (and
their stringy analogs) in the physics of solitons and instantons to string
theory. For example, the string instanton solutions, through vacuum tunneling
computations, may lead to an understanding of the structure of the vacuum in
string theory, much in the same manner as instantons are used in field
theory. The string monopole solutions may appear in grand unification
predictions of string theory, while the macroscopic string solitons may be
used to represent extended string states. An especially noteworthy feature
of the monopoles shown here is the cancellation between gauge and gravitational
singularities in the action, a feature, which, if it survives quantization,
promises to shed light on the nature of string theory as a finite theory of
quantum gravity. A different point of view to these solutions is the study of
the resultant string-inspired low-energy field theories, which may well capture
the essential behaviour of these solitons (e.g. the singularity cancellation
in the monopole solutions) without requiring an expansion in the full string
theory. Finally, there is the open problem of proving the string/fivebrane
duality conjecture$^3$, and its implications to low energy string theory, the
singularity structure of string theory$^4$ and its relation to the more
familiar electric/magnetic duality conjecture of Montonen and Olive$^5$
\vglue 0.6cm
\line{\elevenbf 2. Bosonic and Heterotic Instantons \hfil}
\vglue 0.4cm
We first summarize the 't Hooft ansatz
for the Yang-Mills instanton, and then write down the tree-level bosonic
axionic instanton solution. An exact bosonic solution, with corresponding
CFT can be obtained for the special case of a linear dilaton wormhole.
Finally, an exact multi-instanton solution of heterotic string theory is
obtained, combining the Yang-Mills gauge solution with the bosonic axionic
instanton.

Consider the four-dimensional Euclidean action
$$ S=-{1\over 2g^2}\int d^4x {\rm Tr} F_{\mu\nu}F^{\mu\nu},
\qquad\qquad \mu,\nu =1,2,3,4. \eqno(1)$$
For gauge group $SU(2)$, the fields may be written as $A_\mu=(g/2i)
\sigma^a A_\mu^a$ and $F_{\mu\nu}=(g/2i)\sigma^a F_{\mu\nu}^a$\ \
(where $\sigma^a$, $a=1,2,3$ are the $2\times 2$ Pauli matrices).
A self-dual solution (but not the most general one) to the equation of motion
of this action is given by the 't Hooft ansatz$^6$
$$ A_\mu=i \overline{\Sigma}_{\mu\nu}\partial_\nu \ln f, \eqno(2)$$
where $\overline{\Sigma}_{\mu\nu}=\overline{\eta}^{i\mu\nu}(\sigma^i/2)$
for $i=1,2,3$, where
$$ \eqalign{\overline{\eta}^{i\mu\nu}=-\overline{\eta}^{i\nu\mu}
&=\epsilon^{i\mu\nu},\qquad\qquad \mu,\nu=1,2,3,\cr
&=-\delta^{i\mu},\qquad\qquad \nu=4  \cr} \eqno(3) $$
and where $f^{-1}\Box\ f=0$. The ansatz for the anti-self-dual solution
is similar, with the $\delta$-term in Eq.~(3) changing sign.
To obtain a multi-instanton solution, one solves for $f$ in the
four-dimensional space to obtain
$$ f=1+\sum_{i=1}^N{\rho_i^2\over |\vec x - \vec a_i|^2}, \eqno(4) $$
where $\rho_i^2$ is the instanton scale size and $\vec a_i$ the location in
four-space of the $i$th instanton.

It turns out that there is an analog to the Yang-Mills instanton in the
gravitational sector of the string, namely the axionic instanton$^7$.
In its simplest form, this instanton appears as a solution for the massless
fields of the bosonic string. The bosonic sigma model action can be written as
$$ I={1\over 4\pi\alpha'}\int d^2x\left(\sqrt{\gamma}\gamma^{ab}
\partial_ax^\mu\partial_bx^\nu\met+i\epsilon^{ab}\partial_ax^\mu\partial_b
x^\nu B_{\mu\nu}+\alpha'\sqrt{\gamma}R^{(2)}\phi\right), \eqno(5) $$
where $\met$ is the sigma model metric, $\phi$ the dilaton and $B_{\mu\nu}$
the antisymmetric tensor, and where $\gamma_{ab}$ is the worldsheet metric
and $R^{(2)}$ the two-dimensional curvature. The classical equations of
motion of the effective action of the massless fields is equivalent to
Weyl invariance of $I$. A tree-level solution is given by
any dilaton function satisfying
$e^{-2\phi}\Box\ e^{2\phi}=0$ with
$$ \eqalignno{\met&=e^{2\phi}\delta_{\mu\nu}\qquad \mu,\nu=1,2,3,4,\cr
g_{ab}&=\delta_{ab}\qquad\quad   a,b=5,...,26,\cr
H_{\mu\nu\lambda}&=\pm\epsilon_{\mu\nu\lambda\sigma}\partial^\sigma\phi
\qquad \mu,\nu,\lambda,\sigma=1,2,3,4. &(6) \cr} $$
In order to see the instanton structure of this solution,
we define a generalized curvature $\grijkl$ in terms of the standard
curvature $\rijkl$ and $H_{\mu\alpha\beta}$:
$$ \grijkl=\rijkl+{1\over 2}\left(\nabla_lH^i{}_{jk}-\nabla_kH^i{}_{jl}\right)
+{1\over 4}\left(H^m{}_{jk}H^i{}_{lm}-H^m{}_{jl}H^i{}_{km}\right). \eqno(7) $$
One can also define $\grijkl$ as the Riemann tensor generated
by the generalized Christoffel symbols $\hat\Gamma^\mu_{\alpha\beta}$,
where  $\hat\Gamma^\mu_{\alpha\beta}=\Gamma^\mu_{\alpha\beta}
-(1/2) H^\mu{}_{\alpha\beta}$.
The crucial observation for obtaining higher-loop and even exact solutions
is the following. For any solution given by Eq.~(6),
we can express the generalized curvature in covariant form in terms of
the dilaton field as$^7$
$$ \grijkl=\delta_{il}\nabla_k\nabla_j\phi
-\delta_{ik}\nabla_l\nabla_j\phi+\delta_{jk}\nabla_l\nabla_i\phi
-\delta_{jl}\nabla_k\nabla_i\phi\pm\epsilon_{ijkm}\nabla_l\nabla_m\phi
\mp\epsilon_{ijlm}\nabla_k\nabla_m\phi, \eqno(8) $$
It easily follows that
$$ \grijkl=\mp {1\over 2} \epsilon_{kl}{}^{mn}\hat R^i_{jmn}. \eqno(9) $$
So the instanton appears in the gravitational sector of the string in the
(anti) self-duality of the generalized curvature.
A tree-level multi-instanton solution is therefore
given by Eq.~(6) with the dilaton given by
$$ e^{2\phi}=C+\sum_{i=1}^N {Q_i\over |\vec x -\vec a_i|^2}, \eqno(10) $$
where $Q_i$ is the charge and $\vec a_i$ the location in the four-space
$(1234)$ (the transverse space) of the $i$th instanton.
In the spherically symmetric case $e^{2\phi}={Q/r^2}$, we can explicitly
solve the higher order equations of motion by rescaling the dilaton and
fixing the metric and antisymmetric tensor in lowest order form. For example,
the two-loop dilaton is given by $e^{2\phi}={Q/r^{2(1-{\alpha'\over Q})}}$.
To get an exact solution in this linear dilaton$^8$ case, we
notice that the sigma-model action can be decomposed according to $I=I_1+I_3$,
where for $u=1/r$
$$ I_1={1\over 4\pi\alpha'}\int d^2x\left(Q(\partial u)^2
+\alpha' R^{(2)}\phi\right)\eqno(11) $$
is the action for a Feigin-Fuchs Coulomb gas, a one-dimensional
CFT with central charge given by $c_1=1+6\alpha'(\partial\phi)^2$ and
$I_3$ is the Wess--Zumino--Witten action on an $SU(2)$ group manifold
with central charge
$$ c_3={3k\over k+2}\simeq 3-{6\over k}+{12\over k^2}+...\eqno(12) $$
where $k=Q/\alpha'$, the level of the WZW model, is an integer, from the
quantization condition on the Wess-Zumino term.
Thus $Q$ is not arbitrary, but is quantized in units of $\alpha'$.

We use this splitting to obtain exact expressions
for the fields by fixing the metric and antisymmetric tensor field
in their lowest order form and rescaling the dilaton to all orders in
$\alpha'$. The resulting expression for the dilaton is
$$ e^{2\phi}={Q\over r^{\sqrt{{4\over 1+{2\alpha'\over Q}}}}} \eqno(13) $$
for arbitrary $\alpha'$.

We now turn to the heterotic solution. The tree-level supersymmetric vacuum
equations for the heterotic string are given by
$$ \eqalignno{\delta\psi_M&=\left(\nabla_M-{\textstyle {1\over 4}}H_{MAB}
\Gamma^{AB}\right)\epsilon=0,\cr
\delta\lambda&=\left(\Gamma^A\partial_A\phi-{\textstyle{1\over 6}}
H_{AMC}\Gamma^{ABC}\right)\epsilon=0,\cr
\delta\chi&=F_{AB}\Gamma^{AB}\epsilon=0, &(14) \cr} $$
where $\psi_M,\ \lambda$ and $\chi$ are the gravitino, dilatino and gaugino
fields. The Bianchi identity is given by
$$ dH=\alpha' \left({\rm tr} R\wedge R-{1\over 30}{\rm Tr} F\wedge F\right).
\eqno(15) $$
The $(9+1)$-dimensional Majorana-Weyl fermions decompose down to
chiral spinors according to $SO(9,1)\supset SO(5,1) \otimes SO(4)$ for
the $M^{9,1}\to M^{5,1}\times M^4$ decomposition.
Let $\mu,\nu,\lambda,\sigma=1,2,3,4$ and $a,b=0,5,6,7,8,9$. Then the ansatz
$$ \eqalignno{\met&=e^{2\phi}\delta_{\mu\nu},\cr g_{ab}&=\eta_{ab},\cr
H_{\mu\nu\lambda}&=\pm\epsilon_{\mu\nu\lambda\sigma}\partial^\sigma\phi
&(16) \cr} $$
with constant chiral spinors $\epsilon_\pm$ solves the supersymmetry
equations with zero background fermi fields provided the YM gauge
field satisfies the instanton (anti)self-duality condition
$$ F_{\mu\nu}=\pm {1\over 2}\epsilon_{\mu\nu}{}^{\lambda\sigma}
F_{\lambda\sigma}. \eqno(17) $$
A perturbative solution representing a supersymmetric fivebrane was first
derived by Strominger$^9$. In the absence of a gauge sector, the
multi-fivebrane solution is identical to the tree-level type II supersymmetric
fivebrane solution of Duff and Lu$^{10}$, derived
in terms of the dual seven-form formulation of supergravity,
with $K=e^{-\phi}{}^*H=dA$, where $A$ is the antisymmetric six-form associated
with a fivebrane.
An exact solution is obtained as follows. Define a generalized connection by
$$ \Omega^{AB}_{\pm M}=\omega^{AB}_M\pm H^{AB}_M \eqno(18) $$
embedded in an SU(2) subgroup of the gauge group, and equate it
to the gauge connection $A_\mu~^{11}$ so that $dH=0$ and the corresponding
curvature $R(\Omega_{\pm})$ cancels against the Yang-Mills field strength $F$.
As in the bosonic case, for $e^{-2\phi}\Box\ e^{2\phi}=0$ with the
above ansatz, the curvature of the generalized connection can be written in
covariant form in terms of the dilaton as in Eq.~(8) from which it follows
that both $F$ and $R$ are (anti)self-dual.
This solution becomes exact since $A_\mu=\Omega_{\pm\mu}$
implies that all the higher order corrections vanish$^{7,12,13}$
The self-dual solution for the gauge connection is then given by the 't Hooft
ansatz. An interesting feature of the heterotic
solution is that it combines a YM instanton structure in the gauge
sector with an axionic instanton structure in the gravity sector.
In addition, the heterotic solution has finite action.

Note that the single instanton solution in the heterotic case
carries through to higher order without correction to the dilaton.
This seems to contradict the bosonic solution by suggesting that
the expansion for the central charge $c_3$ terminates at one loop.
This contradiction is resolved by noting that for an $N=4$ worldsheet
supersymmetric solution$^{13}$ the bosonic contribution to the
central charge is given by
$$ c_3={3k'\over k'+2}~,\eqno(19) $$
where $k'=k-2$. This reduces to
$$ c_3=3-{6\over k}=3-{6\alpha'\over Q},\eqno(20) $$
which indeed terminates at one loop order. The exactness of the splitting
then requires that $c_1$ not get any corrections from
$(\partial\Phi)^2$ so that $c_1+c_3=4$ is exact for the tree-level value
of the dilaton$^{7,13}$.
\vglue 0.6cm
\line{\elevenbf 3. Heterotic Multimonopole \hfil}
\vglue 0.4cm
We now turn to a solution with monopole-like structure. We begin with a simple
modification of the 't Hooft ansatz which leads to a multimonopole solution in
field theory, not in the BPS limit. In analogy with the previous section, we
obtain an exact heterotic multimonopole solution$^{14}$.

Going back to the 't Hooft ansatz (Eq.~(1)-(3)), if
we single out a direction in the transverse four-space (say $x_4$)
and assume all fields are independent of this coordinate. Then the solution
for $f$ satisfying $f^{-1} \Box\ f=0$ can be written as
$$ f=1+\sum_{i=1}^N{m_i\over |\vec x - \vec a_i|}, \eqno(21) $$
where $m_i$ is the charge and $\vec a_i$ the location in
the three-space $(123)$ of the $i$th monopole. If we make the identification
$\Phi\equiv A_4$
then the Lagrangian density for the above ansatz can be rewritten as
$$ F_{\mu\nu}^a F_{\mu\nu}^a =F_{ij}^a F_{ij}^a + 2F_{k4}^a F_{k4}^a
=F_{ij}^a F_{ij}^a + 2D_k \Phi^a D_k \Phi^a ,\eqno(22) $$
which has the same form as the Lagrangian density for YM + massless scalar
field in three dimensions.
We now go to $3+1$ dimensions with the Lagrangian density (signature
$(-+++)$)
$$ {\cal L}=-{1\over 4}G_{\mu\nu}^a G^{\mu\nu a} -{1\over 2}
D_\mu \Phi^a D^\mu \Phi^a. \eqno(23) $$
It follows that the above multimonopole ansatz is a static solution with
$A_0^a=0$ and all time derivatives vanish (it is straightforward to verify the
equations of motion). The solution in $3+1$ dimensions has the form
$$ \eqalignno{\Phi^a&=\mp{1\over g}\delta^{ai}\partial_i \omega,\cr
A_k^a&=\epsilon^{akj}\partial_j \omega, &(24)\cr} $$
where $\omega\equiv \ln f$. This solution represents a multimonopole
configuration with sources at $\vec a_i=1,2...N$. A simple observation of
far field and near field behaviour shows that this solution does not arise
in the Prasad-Sommerfield limit. In particular, the fields are singular
near the sources and vanish as $r\to\infty$. This solution can be thought of
as a multi-line source instanton solution, each monopole being interpreted
as an ``instanton string''.

The topological charge of each source is easily computed
($\hat\Phi^a\equiv {\Phi^a/|\Phi|}$) to be
$$ Q=\int d^3x k_0={1\over 8\pi}\int d^3x\epsilon_{ijk}\epsilon^{abc}
\partial_i\hat\Phi^a\partial_j\hat\Phi^b\partial_k\hat\Phi^c=1. \eqno(25) $$
The magnetic charge of each source is then given by $m_i=Q/g=1/g$.
It is also straightforward to show that the Bogomoln'yi bound
$$ G_{ij}^a=\epsilon_{ijk}D_k\Phi^a \eqno(26) $$
is saturated by this solution. Finally, it is easy to show that
the magnetic field $B_i={1\over 2}\epsilon_{ijk}G^{jk}$ (where
$G_{\mu\nu}\equiv \hat\Phi^a F_{\mu\nu}^a-(1/g)\epsilon^{abc}\hat\Phi^a
D_\mu \hat\Phi^b D_\nu \hat\Phi^c$ is the gauge-invariant electromagnetic
field tensor defined by 't Hooft) has the the far field limit behaviour
of a multimonopole configuration:
$$ B(\vec x)\to \sum_{i=1}^N {m_i(\vec x - \vec a_i)\over
|\vec x - \vec a_i|^3},\qquad {\rm as}\quad r\to \infty. \eqno(27) $$
As usual, the existence
of this static multimonopole solution owes to the cancellation of the
gauge and Higgs forces of exchange--the ``zero-force'' condition.

We have presented all the monopole properties of this solution.
Unfortunately, this solution as it stands has divergent
action near each source, and this singularity cannot be simply removed
by a unitary gauge transformation. This can be seen for a single source
by noting that as $r\to 0$, $A_k\to {1\over 2}\left(U^{-1}\partial_k U\right)$,
where $U$ is a unitary $2\times 2$ matrix. The expression in parentheses
represents a pure gauge, and there is no way to get around the $1/2$ factor in
attempting to ``gauge away'' the singularity. The field theory
solution is therefore not very interesting physically.

In analogy with the previous section, we can write down a monopole-like
solution in bosonic string theory and obtain higher-order corrections in the
spherically symmetric case by rescaling the dilaton. However, A CFT
description is elusive in this case as there is no corresponding natural
splitting of the sigma-model action. We turn instead to the heterotic
multimonopole solution. The derivation of this solution closely parallels
that of the heterotic multi-instanton, but
in this case, the solution possesses three-dimensional (rather than
four-dimensional) spherical symmetry near each source. The reduction is
effected by singling out a direction in the transverse space. An exact solution
is now given by
$$ \eqalignno{\met&=e^{2\phi}\delta_{\mu\nu},\qquad g_{ab}=\eta_{ab},\cr
H_{\mu\nu\lambda}&=\pm\epsilon_{\mu\nu\lambda\sigma}\partial^\sigma\phi,\cr
e^{2\phi}&=e^{2\phi_0}f,\cr
A_\mu&=i \overline{\Sigma}_{\mu\nu}\partial_\nu \ln f, &(28)\cr} $$
where in this case $f$ is given by Eq.~(21).
If we again identify the scalar field as $\Phi\equiv A_4$, then the gauge and
scalar fields may be simply written in terms of the dilaton as
$$ \eqalignno{\Phi^a&=-{2\over g}\delta^{ia}\partial_i\phi,\cr
A_k^a&=-{2\over g}\epsilon^{akj}\partial_j\phi &(29)\cr} $$
for the self-dual solution. For the anti-self-dual solution, the scalar
field simply changes sign. Here $g$ is the YM coupling constant of the
heterotic string. Note that $\phi_0$ drops out in Eq.~(29).

The above solution (with the gravitational fields obtained
directly from Eqs.~(21) and (28)) represents an exact multimonopole
solution of heterotic string theory and has the same structure in
the four-dimensional transverse space as the above multimonopole solution
of the YM + scalar field action. If we identify
the $(123)$ subspace of the transverse space as the space part of the
four-dimensional spacetime (with some appropriate toroidal compactification)
and take the timelike direction as the usual $X^0$,
then the monopole properties of the field theory solution carry
over directly into the string solution. Note that the metric line-element
resembles that of the Kaluza-Klein monopole$^{15}$. The string monopole
solution is stable as a result of its saturation of a Bogomol'nyi bound between
ADM mass and charge$^{16}$.

The string action contains a term $-\alpha' F^2$ which also diverges
as in the field theory solution. However, this divergence
is precisely cancelled by the term $\alpha' R^2(\Omega_\pm)$ in the
$O(\alpha')$
action. This result follows from the exactness condition
$A_\mu=\Omega_{\pm\mu}$
which leads to $dH=0$ and the vanishing of all higher order corrections
in $\alpha'$. Another way of seeing this is to consider the higher order
corrections to the bosonic action$^{12}$. All such
terms contain the tensor $T_{MNPQ}$, a generalized curvature incorporating
both $R(\Omega_\pm)$ and $F$. The ansatz is constructed precisely so that this
tensor vanishes identically$^{7,14}$. The action thus reduces to
its finite lowest order form and can be calculated directly for a multi-source
solution from the expressions for the massless fields in the gravity sector.

The divergences in the gravitational sector in heterotic string theory thus
serve to cancel the divergences stemming from the field theory solution. This
solution thus provides an interesting example of how this type of cancellation
can occur in string theory, and supports the promise of string theory as a
finite theory of quantum gravity. Another point of interest is that the string
solution represents a supersymmetric multimonopole solution coupled to gravity,
whose zero-force condition in the gravity sector (cancellation of
the attractive gravitational force and repulsive antisymmetric field force)
arises as a direct result of the zero-force condition in the gauge sector
(cancellation of gauge and Higgs forces of exchange) once the gauge connection
and generalized connection are identified.
\vglue 0.6cm
\line{\elevenbf 4. String Solitons \hfil}
\vglue 0.4cm
In recent work$^{17}$, Dabholkar {\it et al.}
presented a low-energy analysis of macroscopic superstrings and
discovered several interesting analogies between macroscopic superstrings
and solitons in supersymmetric field theories. The main result of this
work centers on the existence of exact multi-string solutions of the
low-energy supergravity super-Yang-Mills equations of motion. In
addition, Dabholkar {\it et al.} find a Bogomolnyi bound for the energy
per unit length which is saturated by these solutions, just as the
Bogomolnyi bound is saturated by magnetic monopole solutions in ordinary
Yang-Mills field theory. The solution may be outlined as follows.
The action for the massless spacetime fields
(graviton, axion and dilaton) in the presence of a source
string can be written as
$$ S = {1\over 2\kappa^2}\int d^Dx\sqrt{g}\left(R-{1\over 2}
{{(\partial\phi)}^2}-{1\over 12}{e^{-2\alpha\phi}}H^2\right)+S_\sigma,
\eqno(30) $$
with the source terms contained in the sigma model action $S_\sigma$ given by
$$ S_\sigma=-{\mu\over 2}\int d^2\sigma(\sqrt{\gamma} \gamma^{mn}
\partial_mX^\mu\partial_nX^\nu g_{\mu\nu}e^{\alpha\phi} +
\epsilon^{mn}\partial_mX^\mu\partial_nX^\nu B_{\mu\nu}), \eqno(31) $$
with $\alpha=\sqrt{2/(D-2)}$ and $\gamma_{mn}$ a worldsheet metric to be
determined.
The sigma model action $S_\sigma$ describes the coupling of the string to the
metric, antisymmetric tensor field and dilaton. The first part of the
action $S$ above represents the effective action for the
massless fields in the spacetime frame and
whose equations of motion are equivalent to conformal invariance of the
underlying sigma model. The combined action thus generates the equations
of motion satisfied by the massless fields in the presence of a
macroscopic string source.
The static solution to the equations of motion is given by
$$ \eqalignno{ds^2&=e^{A}[-dt^2+(dx^1)^2]+e^{B}d\vec x \cdot d\vec x\cr
A&={D-4\over D-2}E(r)\qquad B=-{2\over D-2}E(r)\cr
\phi&=\alpha E(r)\qquad\qquad B_{01}=-e^{E(r)},&(32)\cr} $$
where $x^1$ is the direction along the string, $r=\sqrt{\vec x \cdot\vec x}$
and
$$ e^{-E(r)}=\cases{1+{M\over r^{D-4}}&$D>4$\cr 1-8G\mu\ln(r)&$D=4$\cr} $$
for a single static string source. The solution can be generalized to
an arbitrary number of static string sources by linear superposition of
solutions of the ($D-2$)-dimensional Laplace's equation. A similar solution
incorporating the eight-dimensional instanton of Grossman, Kephart and
Stasheff in the gauge sector was derived by Duff and Lu$^{18}$.

The force exerted on a test string moving in given background fields
is obtained from the sigma model equation of motion
$$ \nabla_m(\gamma^{mn}\nabla_nX^\mu)=-\Gamma^\mu_{\nu\rho}
\partial_mX^\nu\partial_nX^\rho\gamma^{mn}+{1\over 2}
H^\mu{}_{\nu\rho}\partial_mX^\nu\partial_nX^\rho\epsilon^{mn},\eqno(33) $$
where $\Gamma^\mu_{\nu\rho}$ are the Christoffel symbols calculated from
the sigma model metric $G_{\mu\nu}=g_{\mu\nu}e^{\alpha\phi}$. We make the
usual distinction between the sigma model metric and the Einstein metric;
spacetime indices are raised and lowered by contraction with $G_{\mu\nu}$;
worldsheet indices are denoted by $m$ and $n$.
Consider a stationary test string in the background of a source
string located at the origin. Assume further that both strings run along
the $x_1$ direction and have the same orientation. We use conformal
gauge for the test string and get $X^0=\tau$, $X^1=\sigma$,
$\gamma_{mn}=diag(-1,+1)$ and $\epsilon^{01}=+1$. The transverse force
then vanishes
$$ {d^2\over d\tau^2} X^i=-2\Gamma^i_{00}+H^i{}_{10}=0. \eqno(34) $$
Note that if the test string and source string were oppositely oriented
then the second term would appear with a negative sign and there would
be a net attractive force. Also note that the no-force condition depends only
on the general ansatz and not on the precise form of the solution.

The zero-force condition arises from the cancellation of
long-range forces of exchange of the massless fields of the string
(the graviton, axion and dilaton) and can be seen explicitly from
Eq.~(34). This is a perfect analog to the zero-force condition of
Manton for magnetic monopoles, which requires that the attractive
scalar exchange force precisely cancel the repulsive vector exchange
force when the Bogomolnyi bound is attained. Dabholkar {\it et al.}
show that a similar Bogomolnyi bound is satisfied by their string
soliton solutions, further strengthening the analogy with the
monopoles. In the next section, however, we shall see that in contrast
to BPS monopoles, the string solitons also obey a zero {\it dynamical}
force condition.
\vglue 0.6cm
\line{\elevenbf 5. Dynamics \hfil}
\vglue 0.4cm
We now summarize some recent results on the dynamics of these solitons$^2$.
While the static
force vanishes as a result of the cancellation of long-range forces of
exchange, the force between two moving solitons is in general
nonvanishing and depends on the velocities of the solitons. The most
complete answer would be given by a full time-dependent solution of the
equations of motion of the above action for the case of an arbitrary
number of sources moving with arbitrary transverse velocities. These
equations, however, are much more difficult to solve for moving sources
than for a static configuration. Even a two-soliton
solution is in general quite intractable for this class of actions.

We first examine the scattering of these solitons using the above test-string
approach. This entails solving the constraint equation for the
worldsheet metric obtained by varying the worldsheet Lagrangian ${\cal
L}$. The resultant solution for the worldsheet metric along with the
static solution for the spacetime metric, antisymmetric tensor field
and dilaton from the static ansatz for a single source string are then
substituted into the Lagrangian, whose equations yield the dynamics of
the test string in the source string background. A flat kinetic Lagrangian
is obtained, suggesting trivial scattering (i.e. a zero dynamical force) in
this limit.

We then address the scattering problem from a string-theoretic point of
view. The winding configuration described by $X(\sigma,\tau)$ describes a
soliton string state. It is therefore a natural choice for us to compare
the dynamics of these states with the above string solitons in order to
determine whether we can identify these solitons with infinitely long
fundamental strings. Accordingly, we study the scattering of the winding states
in the limit of large winding radius. We find that the Veneziano amplitude
obtained also indicates trivial scattering in this limit, providing
evidence for the identification of the string solitons with infinitely
long macroscopic fundamental strings.

Finally, we turn to soliton-soliton scattering.
In the low-velocity limit, multi-soliton solutions trace out geodesics in
the static solution manifold, with distance defined by the Manton metric on
moduli space manifold. In the absence of a full time-dependent
solution to the equations of motion, these geodesics represent a good
approximation to the low-energy dynamics of the solitons. For BPS monopoles,
the Manton procedure was implemented by Atiyah and Hitchin$^{19}$.
Computing the Manton metric on moduli space for the scattering of the soliton
string solutions in $D=4$ (we expect that the same result will hold for
arbitrary $D \geq 4$) we find that the metric is flat to lowest
nontrivial order in the string tension. This result implies trivial
scattering of the string solitons and is consistent with the above two
calculations, and thus provides even more compelling evidence for the
identification of the string soliton with the underlying fundamental string.

For the instanton and monopole fivebranes$^{20}$, both the test-fivebrane limit
and the metric on moduli space also yield a zero dynamical force condition.
Since a fundamental theory of fivebranes has not yet been constructed,
there is no corresponding Veneziano amplitude computation with which to
compare. For the instantons, it is sufficient to demonstrate Ricci flatness of
the Manton metric to obtain trivial scattering while for the monopoles a flat
metric can be explicitly computed. The zero dynamical force can be seen as a
direct consequence of the exactness condition of equating the gauge and spin
connections$^{16}$.
\vglue 0.6cm
\line{\elevenbf 6. Future Directions \hfil}
\vglue 0.4cm
While all the solutions we discussed are classical, one can still conceive
of situations in which quantum corrections to the instantons, for
example, drop out in nonperturbative computations (such as for vacuum
tunnelling). To this end, a vertex operator representation of the instantons
would be highly desirable. The most interesting feature of the heterotic
monopole solution is the cancellation between gauge and gravitational
singularities. If this is an intrinsically ``stringy'' feature, then
it presumably occurs in a larger context within string theory, in which case
the full quantum
string loop extension of this solution promises to shed light on the nature
of string theory as a finite theory of quantum gravity. If this cancellation
is of a more accidental nature, then it would pay to concentrate more on the
corresponding low-energy field theory, whose quantization is presumably far
simpler. In either case, it would be interesting to see whether the
singularity cancellation occurs in a quantized solution, or in the context
of blackhole type solutions. We have compelling dynamical evidence for the
identification of the string solitons with macroscopic
fundamental strings, but an exact heterotic solution seems most natural
in the context of the conjectured dual fundamental theory of fivebranes.
While the construction of such a theory remains elusive, there is so far
solid evidence to support the duality conjecture (see M.~J.~Duff's
contribution to this volume).
\vglue 0.6cm
\line{\elevenbf Acknowledgements \hfil}
\vglue 0.4cm
It is a great pleasure to thank Mike Duff for helpful discussions, fruitful
collaboration and general advice and encouragement over the past eighteen
months. I am very grateful to the World Laboratory for giving me the
means to pursue my work. This work was also supported in part by the
National Science Foundation.
\vglue 0.6cm
\line{\elevenbf References \hfil}
\vglue 0.4cm
\medskip
\item{1.} R.~R.~Khuri, {\elevenit Phys. Rev.} {\elevenbf D46} (1992) 4526.
\item{2.} R.~R.~Khuri, {\elevenit Geodesic Scattering of Solitonic Strings},
Texas A\&M preprint CTP/TAMU-79/92; {\elevenit Classical Dynamics of
Macroscopic Strings}, Texas A\&M preprint CTP/TAMU-80/92 (to appear in
Nucl. Phys. B).
\item{3.} M.~J.~Duff, {\elevenit Class. Quan. Grav.} {\elevenbf 5} (1988)
189; M.~J.~Duff and J.~X.~Lu, {\elevenit Nucl. Phys.} {\elevenbf B354} (1991)
129.
\item{4.} M.~J.~Duff, R.~R.~Khuri and J.~X.~Lu, {\elevenit Nucl. Phys.}
{\elevenbf B377} (1992) 281.
\item{5.} C.~Montonen and D.~Olive, {\elevenit Phys. Lett.} {\elevenbf B72}
(1977) 117; H.~Osborn, {\elevenit Phys. Lett.} {\elevenbf B83} (1979) 321.
\item{6.} G.~'t~Hooft, {\elevenit Nucl. Phys.} {\elevenbf B79} (1974) 276;
{\elevenit Phys. Rev. Lett.} {\elevenbf 37} (1976) 8;
F.~Wilczek, in {\elevenit Quark confinement and field theory},
ed. D.~Stump and D.~Weingarten, (John Wiley and Sons, New York,1977);
E.~Corrigan and D.~B.~Fairlie, {\elevenit Phys. Lett.} {\elevenbf B67} (1977)
69; R.~Jackiw, C.~Nohl and C.~Rebbi, {\elevenit Phys. Rev.} {\elevenbf D15}
(1977) 1642; {\elevenit Particles and Fields}, ed. David Boal and
A.~N.~Kamal, (Plenum Publishing Co., New York,1978) p.199.
\item{7.} R.~R.~Khuri, {\elevenit Phys. Lett.} {\elevenbf B259} (1991) 261;
{\elevenit Proceedings of XXth International Conference on
Differential Geometric Methods in Theoretical Physics}, ed. S.~Catto and
A.~Rocha (World Scientific, Jan. 1992) p.1074.
\item{8.} I.~Antoniadis, C.~Bachas, J.~Ellis and D.~V.~Nanopoulos,
{\elevenit Phys. Lett.} {\elevenbf B211} (1988) 393;
{\elevenit Nucl. Phys.} {\elevenbf B328} (1989) 117.
\item{9.} A.~Strominger, {\elevenit Nucl. Phys.} {\elevenbf B343} (1990) 167.
\item{10.} M.~J.~Duff and J.~X.~Lu, {\elevenit Nucl. Phys.} {\elevenbf B354}
(1991) 141.
\item{11.} J.~M.~Charap and M.~J.~Duff, {\elevenit Phys. Lett.}
{\elevenbf B69} (1977) 445.
\item{12.} E.~A.~Bergshoeff and M.~de Roo, {\elevenit Nucl. Phys.}
{\elevenbf B328} (1989) 439; {\elevenit Phys. Lett.} {\elevenbf B218}
(1989) 210.
\item{13.} C.~G.~Callan, J.~A.~Harvey and A.~Strominger, {\elevenit Nucl.
Phys.} {\elevenbf B359} (1991) 611.
\item{14.} R.~R.~Khuri, {\elevenit Phys. Lett.} {\elevenbf B294} (1992) 325;
{\elevenit Nucl. Phys.} {\elevenbf B387} (1992) 315.
\item{15.} D.~J.~Gross and M.~J.~Perry, {\elevenit Nucl. Phys.}
{\elevenbf B226} (1983) 29; R.~D.~Sorkin, {\elevenit Phys. Rev. Lett}
{\elevenbf 51} (1983) 87.
\item{16.} R.~R.~Khuri, {\elevenit A Comment on the Stability of String
Monopoles}, Texas A\&M preprint CTP/TAMU-81/92.
\item{17.} A.~Dabholkar and J.~A.~Harvey, {\elevenit Phys. Rev. Lett.}
{\elevenbf 63} (1989) 719; A.~Dabholkar, G.~Gibbons, J.~A.~Harvey and
F.~Ruiz Ruiz, {\elevenit Nucl. Phys.} {\elevenbf B340} (1990) 33.
\item{18.} B.~Grossman, T.~W.~Kephart and J.~D.~Stasheff, {\elevenit Commun.
Math. Phys.} {\elevenbf 96} (1984) 431; {\elevenit Commun. Math. Phys.}
{\elevenbf 100} (1985) 311; {\elevenit Phys. Lett.} {\elevenbf B220} (1989)
431; M.~J.~Duff and J.~X.~Lu, {\elevenit Phys. Rev. Lett.} {\elevenbf 66}
(1991) 1402.
\item{19.} N.~S.~Manton, {\elevenit Phys. Lett.} {\elevenbf B110} (1982) 54;
M.~F.~Atiyah and N.~J.~Hitchin, {\elevenit Phys. Lett.} {\elevenbf A107}
(1985) 21; {\elevenit The Geometry and Dynamics of Magnetic Monopoles},
(Princeton University Press,1988).
\item{20.} R.~R.~Khuri, {\elevenit Nucl. Phys.} {\elevenbf B376} (1992) 350;
C.~G.~Callan and R.~R.~Khuri, {\elevenit Phys. Lett.} {\elevenbf B261}
(1991) 363; R.~R.~Khuri, {\elevenit Phys. Lett.} {\elevenbf B294} (1992)
331; A.~G.~Felce and T.~M.~Samols, {\elevenit Low-Energy Dynamics of String
Solitons} ITP Santa Barbara preprint NSF-ITP-92-155.

\eject
\bye